# A STUDY OF A DETERMINISTIC MODEL FOR MENINGITIS EPIDEMIC

**S. J Yaga**
On Study leave (Ph.D.)
Department of Statistics,
University of Abuja,
Abuja
Nigeria.
samailajyaga@unimaid.edu.ng

**F .W. O Saporu**
Visiting Professor
Department of Statistics,
University of Abuja,
Abuja,
Nigeria.
saporuf@gmail.com

**ABSTRACT**

A compartmental deterministic model that allows (1) immunity from two stages of infection and carriage, and (2) disease induced death, is used in studying the dynamics of meningitis epidemic process in a closed population. It allows for difference in the transmission rate of infection to a susceptible by a carrier and an infective. It is generalized to allow a proportion ($\varphi$) of those susceptibles infected to progress directly to infectives in stage I. Both models are used in this study. The threshold conditions for the spread of carrier and infectives in stage I are derived for the two models. Sensitivity analysis is performed on the reproductive number derived from the next generation matrix. The case-carrier ratio profile for various parameters and threshold values are shown. So also are the graphs of the total number ever infected as influenced by $\epsilon$ and $\varphi$. The infection transmission rate ($\beta$), the odds in favor of a carrier, over an infective, in transmitting an infection to a susceptible ($\epsilon$) and the carrier conversion rate ($\phi$) to an infective in stage I, are identified as key parameters that should be subject of attention for any control intervention strategy. The case-carrier ratio profiles provide evidence of a critical case-carrier ratio attained before the number of reported cases grows to an epidemic level. They also provide visual evidence of epidemiological context, in this case, epidemic incidence (in later part of dry season) and endemic incidence (during rainy season). Results from total proportion ever infected suggest that the model, in which $\varphi = 0$ obtained, can adequately represent, in essence, the generalized model for this study.

**Correspondence Address:** P.O.BOX 1069, University of Maiduguri, Department of Mathematical Sciences, Maiduguri, Borno state, Nigeria. **Email:** samailajyaga@unimaid.edu.ng

# 1. Introduction

Meningitis is the infection of the meninges, a thin lining covering the brain and the spinal cord. The disease can be caused by many different pathogens (bacteria, fungi, or viruses). The highest global burden is seen with bacteria (meningococcus) meningitis (WHO, 2021). Meningococcus is a pathogenic member of the Neisseria genus. The human pathogens are Neisseria meningitidis (N.meningitidis) and N.gonorrhea or gonococcus. Meningococcus can be divided into thirteen serogroups with six (A, B, C, W-135, X and Y) linked with pathogenic ability (Ahmed-Abubakar, 2014 and Batista et al., 2017).

The common manifestations of meningococcal disease are meningitis, where the bacteria are found in the cerebrospinal fluid and septicemia, where the bacteria are found in the blood (Raman, 1988). Meningococcal disease occurs worldwide as an endemic disease with seasonal fluctuations (Stephens et al., 2007 and Caugant et al., 2012). The epidemic pattern is concentrated mainly in the African meningitis belt, a region that spans the continent from Senegal to Ethiopia (Lapeyssonie et al., 1963; Molesworth et al., 2002). It is a devastating disease and remains a major public health challenge.

# 2. Literature Review

Mathematical models play key role in gaining proper insight in the understanding of the principles that govern the transmission dynamics of diseases. Deterministic compartmental models (Saporu, 1987; Zhao et al., 2000; and Alemneh and Belay, 2022) have been mostly used. The first model (Coen et al., 2000) is a Susceptible, Carrier (infected), and Recovered (SCR) model applied to age prevalence data. An endemic model was used by Vereen (2008) for analyzing the impact of vaccination program. Some age-structured dynamic models for vaccination program can be found in Cardine et al. (2005), Karachaliu et al. (2015) and Asamoah et al. (2018). Models that allow for seasonality of meningitis can be found in, for examples, Agier et al. (2013) and Irving et al. (2012).

# 3. The Problem

In Irving et al. (2012) four endemic models were considered, SCIS, SCIRS$^I$, SCIRS$^{CI}$ and SCIRS$^{ALT}$ for an endemic situation, allowing natural forces of birth and death. Here interest lies in two infective states; I, Infective (without complications) and $I_1$ infective (with complications). We also allow for death from illness only. The focus here is to use this model in an epidemic situation of a closed population, in order to bring about a better understanding of the transmission dynamics of meningitis epidemic process.

## 4. Meningitis

Susceptible individuals acquire pathogen after exposure through effective and prolonged contact with asymptomatic carriers or infectious individuals (Meyer and Kristiansen, 2016). Asymptomatic carriers harbor the pathogen in the nasopharynx and spread through respiratory secretions. Carriers may lose carriage naturally (Trotter and Maiden, 2016) or develop invasive disease (Irving et al., 2012) with duration of carriage between 5 to 6 months (3.4 months on average in Africa meningitis belt). Infected individuals undergo an incubation period of 2 to 10 days (WHO, 2021) before becoming infectious. The symptoms associated with the disease include headache, stiff neck, fever, etc. Within 3 to 7 days after exposure, infectious individuals progress to stage I complication where they can experience organ failure, seizures, circulatory collapse, etc. If untreated, they progresses to stage II complications where they may experience blindness, memory loss, gangrene leading to amputation (WHO, 2021; Sharew et al., 2020). With early detection, individuals with early infection may recover with no complication after treatment with antibiotics for 7 days. Stage I complication individuals can recover with stage I complications and individuals with stage II complications recover with stage II complication after treatment, or die within 14 days of admission (MacMillan et al.,2001). Recovery does not confer lifelong immunity.

Meningitis occurs during the dry season, usually mid-October and dies out with the onset of rainfall in mid-April. Carriers sustain the pathogen during raining season. These cause later epidemics. The ratio of carrier to cases is likely to be greater than 100:1 (Greenwood, 2006).

The risk of meningitis varies with age. Carrier prevalence is also age dependent (Campagne et al., 1999). Seasonality is a factor that also influences meningitis incidence.

## 5. The Model

Mathematical models are developed and used in order to gain insight into the underlying mechanism of a disease epidemic process. Here we consider a compartmental model that allows important features that can adequately represent the dynamics of meningococcal infection. In order to make the model simple and mathematically tractable, we consider only six mutually exclusive epidemiological classes (states) for classifying an individual. In a meningitis epidemic process, an individual can be susceptible ($S$), asymptomatic carrier($C$), symptomatic infectious individual or ill ($I$), infectious individual with complications ($I_1$), recovery/immuned (R) or dead (D) due only to illness of the disease. Notice here that we allow only two infectious classes instead of three and only one recovered class, $R$, where recovered individuals from any of the two stages of infectiousness are placed to make the model simple and tractable. Death from illness is allowed only after an individual has progressed to the complicated stage and fails to recover. A susceptible individual can be infected either by an asymptomatic carrier ($C$) or an infectious individual ($I$) with transmission rate $\beta$. The carrier-case ratio is known (Irving et al., 2011) to be above 100:1. This gives credence to a differential transmission rate. In order to reflect this, we assume a force of infection

$$\lambda = \frac{\beta(\epsilon C + I)}{N}, \qquad (1)$$

where, $N$ is the total population size and $\epsilon$ is the odds in favor of a carrier transmitting infection to a susceptible. Notice here that $I_1$ is not in equation (1). This is because it is reasonable to assume that an infective in stage $I_1$ will not be out for circulation due to confinement arising from the complications of the infection. All infected individuals must pass through carriership before becoming an infective in stage I.

In order to generalize our model, we allow a proportion $\varphi$ of infectives to pass directly from $S$ to $I$. Carriers can either develop invasive disease ($I$) at the rate $\phi$ or lose carriage at the rate $\sigma$ to become susceptible again. An infective in state $I$ either progresses to state $I_1$ at the rate $\theta$ or recovers at the rate $\gamma_1$. An infective in state $I_1$ either dies at the rate $\delta$ or recovers at the rate $\gamma_2$. Recovery confers only temporary immunity so that a recovered individual becomes susceptible again at the rate $\alpha$.

The model assumptions are close to birth, natural death and migration (in or out). Clearly, this model is an extension of SCIRS$^{ALT}$ model of Irving et al. (2012), in a closed population. The model schematic representation is given in Fig.1 below.

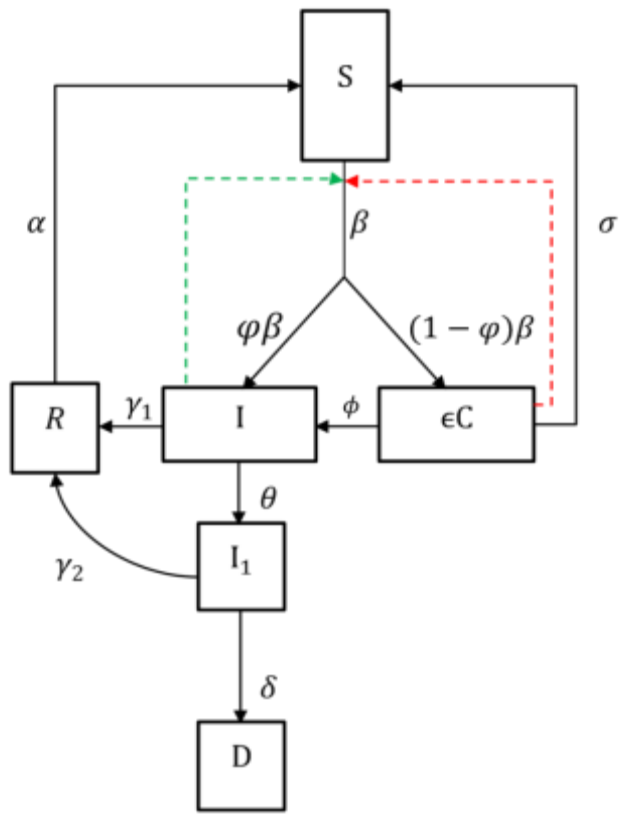

**Fig 5.1** Schematic representation of the meningitis epidemic model

For the ease of understanding, we present two systems of model equations below.

Model 1 : $\varphi \neq 0$

$$\begin{aligned}
\frac{dS}{dt} &= -\frac{\beta S(\epsilon C + I)}{N} + \sigma C + \alpha R \\
\frac{dC}{dt} &= \frac{(1-\varphi)\beta S(\epsilon C + I)}{N} - (\phi + \sigma)C \\
\frac{dI}{dt} &= \frac{\varphi \beta S(\epsilon C + I)}{N} + \phi C - (\theta + \gamma_1)I \\
\frac{dI_1}{dt} &= \theta I - (\delta + \gamma_2)I_1 \\
\frac{dR}{dt} &= \gamma_1 I + \gamma_2 I_1 - \alpha R \\
\frac{dD}{dt} &= \delta I_1
\end{aligned} \qquad (2)$$

where

$$S(t) + C(t) + I(t) + I_1(t) + R(t) + D(t) = N.$$

The model equation for proportion is obtained by dividing the number of individuals in all compartments at time $t$ by the total population $N$ to get:

$$\begin{aligned}
\frac{ds}{dt} &= -\frac{\beta s(\epsilon c + i)}{N} + \sigma c + \alpha r \\
\frac{dc}{dt} &= \frac{(1-\varphi)\beta s(\epsilon c + i)}{N} - (\phi + \sigma)c \\
\frac{di}{dt} &= \frac{\varphi \beta s(\epsilon c + i)}{N} + \phi c - (\theta + \gamma_1)i \\
\frac{di_1}{dt} &= \theta i - (\delta + \gamma_2)i_1 \\
\frac{dr}{dt} &= \gamma_1 i + \gamma_2 i_1 - \alpha r \\
\frac{dz}{dt} &= \delta i_1
\end{aligned} \qquad (3)$$

where,

$$s(t) + c(t) + i(t) + i_1(t) + r(t) + z(t) = 1.$$

Model 2: $\varphi = 0$

$$\begin{aligned}
\frac{dS}{dt} &= -\frac{\beta S(\epsilon C + I)}{N} + \sigma C + \alpha R \\
\frac{dC}{dt} &= \frac{\beta S(\epsilon C + I)}{N} - (\phi + \sigma)C \\
\frac{dI}{dt} &= \phi C - (\theta + \gamma_1)I \\
\frac{dI_1}{dt} &= \theta I - (\delta + \gamma_2)I_1 \\
\frac{dR}{dt} &= \gamma_1 I + \gamma_2 I_1 - \alpha R \\
\frac{dD}{dt} &= \delta I_1
\end{aligned} \right\} \quad (4)$$

where,

$$S(t) + C(t) + I(t) + I_1(t) + R(t) + D(t) = N,$$

and the system of equation for proportion is given by

$$\begin{aligned}
\frac{ds}{dt} &= -\frac{\beta s(\epsilon c + i)}{N} + \sigma c + \alpha r \\
\frac{dc}{dt} &= \frac{\beta s(\epsilon c + i)}{N} - (\phi + \sigma)c \\
\frac{di}{dt} &= \phi c - (\theta + \gamma_1)i \\
\frac{di_1}{dt} &= \theta i - (\delta + \gamma_2)i_1 \\
\frac{dr}{dt} &= \gamma_1 i + \gamma_2 i_1 - \alpha r \\
\frac{dz}{dt} &= \delta i_1
\end{aligned} \right\} \quad (5)$$

where,

$$s(t) + c(t) + i(t) + i_1(t) + r(t) + z(t) = 1.$$

The description of model parameters and reasonable intuitive estimates are given in Table 1 below.

**Table 1.** Description of parameter values used in the model

| Parameters | Interpretations | Values/Ranges /day | Source |
|---|---|---|---|
| $\epsilon$ | Increased infectivity rate of carriers ($C$) | 0.5-10 | Assumed |
| $\beta$ | Transmission rate | 0.13699-0.54795 | Irving et al. (2012) |
| $\varphi$ | Proportion of $S$ moving directly to $I$ | 0-1 | Assumed |
| $\sigma$ | rate at which individuals in $C$ moving back to $S$ | 0.002739-0.14247 | Irving et al. (2012) |
| $\phi$ | rate at which individuals $C$ moves to $I$ | 0.002739-0.14247 | Irving et al. (2012) |
| $\theta$ | rate at which individuals in $I$ moves to $I_1$ | 0.071428-0.14247 | WHO (2018) |
| $\gamma_1$ | rate at which individuals in $I$ recover | 0.1-0.14247 | Irving et al. (2012) |
| $\gamma_2$ | rate at which individuals in $I_1$ recover | 0.08333-0.1 | Assumed |
| $\alpha$ | Rate at which individuals in $R$ moves to $S$ | 0.0333 | Irving et al. (2012) |
| $\delta$ | Rate at which individuals in $I_1$ die | 0.08333-0.1 | Sharew et al. (2020) |
| **Variables** | | | |
| S | Susceptible | | |
| C | Carrier | | |
| I | Stage $I$ infective with no complication | | |
| $I_1$ | Stage $I_1$ infective with Complication | | |
| R | Recovery/Immune | | |
| D | Death due to disease | | |

**Parameter values are assumed within realistic ranges after discussion with medical doctors and professionals in the field.**

Model plots for various parameter values are given in Fig 2 below.

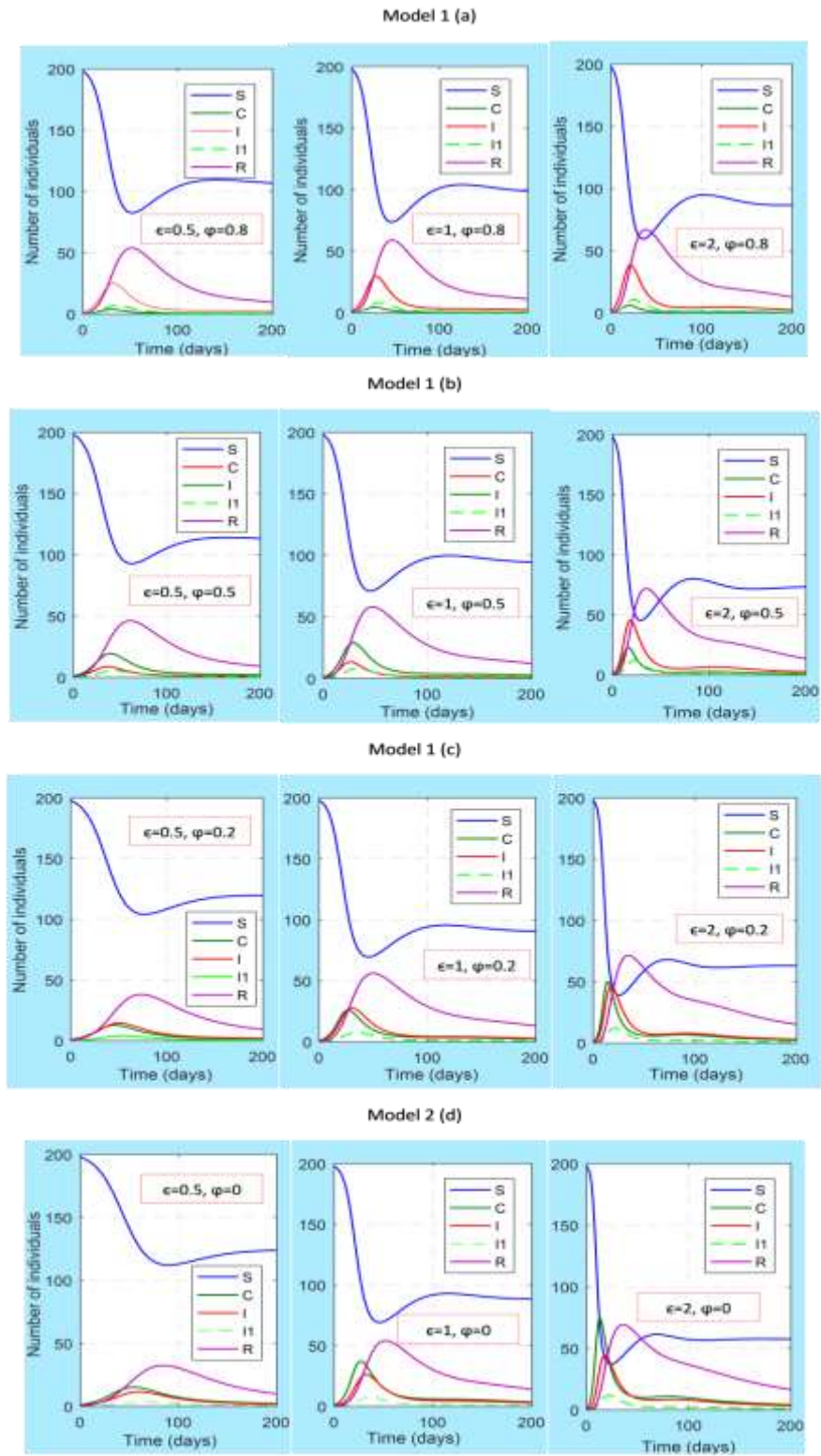

**Fig 5.2**. model plots for varying parameter values $\epsilon$ and $\varphi$ for initial conditions I(0)=1, C(0)=1, for model 1: (a) $\epsilon = 0.5, 1, 2; \varphi = 0.8$ (b) $\epsilon = 0.5, 1, 2; \varphi = 0.5$ (c) $\epsilon = 0.5, 1, 2; \varphi = 0.2$ and for model 2 (d) $\epsilon = 0.5, 1, 2; \varphi = 0$

# 6. Threshold Properties of Models

Asymptomatic carriers ($C$) and infectives ($I$) are individuals primarily responsible for the spread of meningococcal infection. In order to understand the dynamics of the spread of this disease, therefore, we investigate the conditions for the initial spread of each of these infectious stages in the two models described above.

**Model I**

**A. Condition for initial spread of carriers**

From equation (2)

$$\left((1-\varphi)\beta\frac{S(0)}{N}\right)I(0) + \left(\epsilon(1-\varphi)\beta\frac{S(0)}{N} - (\phi+\sigma)\right)C(0) > 0. \tag{6}$$

This implies that if there are no initial number of carriers (that is $C(0) = 0$), there will still be the spread of carriers, provided, $\varphi \neq 1$ and the initial number of infectives $I(0) \neq 0$. This is as expected because some individuals in stage I will produce carriers by infecting susceptibles. However, if $I(0) = 0$ and $C(0) \neq 0$, the number of carriers will grow provided,

$$\frac{S(0)}{N} > \frac{\phi+\sigma}{\epsilon(1-\varphi)\beta}. \tag{7}$$

Consequently, the number of carriers will grow provided the initial proportion of susceptibles $S(0)/N$ exceeds a threshold value

$$\rho_C = \frac{\phi+\sigma}{\epsilon(1-\varphi)\beta}, \tag{8}$$

whenever, the initial number of infectives is zero. We can also interpret this threshold property by allowing $S(0) \cong N$ in equation (4) and it becomes $_cR_0^1 > 1$ where

$$_cR_0^1 = \frac{\epsilon(1-\varphi)\beta}{\phi+\sigma}. \tag{9}$$

In order to be able to interprete $_cR_0^1$, we assume exponential times between infective contacts by a carrier and denote the average time between these contacts by

$$T_C = \frac{1}{\epsilon(1-\varphi)\beta}. \tag{10}$$

Similarly, the average stay in the carrier stage is denoted by

$$T_{SC} = \frac{1}{\phi + \sigma}. \tag{11}$$

Hence the number of secondary infectives produced by a carrier is

$$\frac{T_{SC}}{T_C} = \frac{\epsilon(1-\varphi)\beta}{\phi + \sigma} = {}_cR_0^1, \tag{12}$$

which is the infective reproductive ratio produced by a carrier when the initial number of infectives is zero in a population that is entirely susceptible. Consequently, the threshold property derived from equation (7) can equally be stated as follows.

**Threshold property I**

In a closed population when the initial number of infectives is zero, the number of meningitis carrier can only grow if there is a threshold fraction of susceptible exceeding $1/{}_cR_0^1$ or, if ${}_cR_0^1 > 1$.

In this circumstance, if control intervention is targeted at reducing the proportion of susceptibles below $1/{}_cR_0^1$ meningitis can be eradicated.

The reader should note the dual epidemiological interpretation of ${}_cR_0^1$. It gives the infective reproduction number produced by carriers and provides the condition for the initial spread of carriers.

**B. Condition for initial spread of infectives**

From equation (2),

$$\left(\frac{\varphi\beta}{\theta + \gamma_1}\left(\frac{S(0)}{N}\right) - 1\right)I(0) + \left(\frac{\epsilon\varphi\beta}{\theta + \gamma_1}\left(\frac{S(0)}{N}\right) + \frac{\phi}{\theta + \gamma_1}\right)C(0) > 0. \tag{13}$$

It is clear that when $I(0) = 0$, infection will spread provided $C(0) > 1$. If $C(0) = 0$ and $I(0) \neq 0$

$$\frac{S(0)}{N} > \frac{\theta + \gamma_1}{\varphi\beta} = \rho_I. \tag{14}$$

This provides a threshold value for the initial proportion of susceptibles different from equation (4). By similar argument as before we define

$${}_IR_0^1 = \frac{\varphi\beta}{\theta + \gamma_1}, \tag{15}$$

as the number of secondary infectives produced by a case that did not go through carriership. Hence this threshold property can equally be stated as follows.

**Threshold property II**

In a closed population, when the initial number of carriers is zero, the number of meningitis cases can only grow if there is a threshold fraction of susceptibles exceeding $1/{}_IR_0^1$ or, if ${}_IR_0^1 > 1$.

In this circumstance, if a control intervention is target at reducing the proportion of susceptibles below $1/{}_IR_0^1$, meningitis can be eradicated. Considering both threshold results, that is, ${}_cR_0 > 1$ and ${}_IR_0^1 > 1$, a control intervention targeted at reducing the proportion of susceptibles below the inverse of whichever is greater of these two reproductive ratios will cause the eradication of meningitis. Equivalently, keeping the smaller of these ratios ( ${}_IR_0^1$ and ${}_cR_0^1$) less than one will cause the eradication of meningitis.

**Model 2**

**A. Condition for the spread of Carrier**

Model 2 is a special case of model 1 when $\varphi = 0$. Here from equation (4)

$$_cR_0^2 = \frac{\epsilon\beta}{(\phi+\sigma)}. \tag{16}$$

Conclusions about the threshold properties similarly follow.

**B. Condition for the spread of Infectives**

From equation (4), the condition for the spread of infectives is given by

$$\phi C(0) - (\theta + \gamma_1)I(0) > 0. \tag{17}$$

If initially $I(0) = 0$ and $C(0) \neq 0$, infection will definitely spread. If $C(0) = 0$ and $I(0) \neq 0$ infection cannot spread. Consequently, $C(0) \neq 0$ is crucial to the spread of infection. There is a scenario when $C(0) \neq 0$ and $I(0) \neq 0$. In this case for infection to spread

$$\frac{\phi}{(\theta + \gamma_1)} > \frac{I(0)}{C(0)}. \tag{18}$$

This provides a critical case-carrier ratio threshold condition for infection to spread. Denote this by

$$\rho_{ccc} = \frac{\phi}{(\theta + \gamma_1)}. \tag{19}$$

By interpretation $\rho_{CCC}$ is the number of carriers that become infectives during the average infectious period of an infective. Consequently, the threshold result depicted in equation (18) can be expressed as follows.

**Threshold property III**

In a closed population, where the initial number of cases and carriers are not zero, meningitis will spread whenever the critical case-carrier ratio, $\rho_{CCC}$, is greater than the ratio of the number of cases to carriers.

### C. Case Carrier Ratio Trajectory

Case-carrier ratio is an ecological proxy (Kountangni et al., 2015) for the risk of meningitis given colonization. Consequently a case-carrier ratio trajectory, over time will provide visual evidence on how meningitis incidence varies according to epidemiological context (endemicity, hyperendemicity and epidemic) times. This will bring about a better understanding of meningitis transmission process that will aid in designing control strategies, including vaccination programmes. The case-carrier ratio threshold result derived earlier enables us to write

$$\frac{I(t)}{C(t)} < \rho_{ccc}. \tag{20}$$

In order to study this trajectory, solving this equation for time, $t$, is difficult. However, an approximate graphical solution is feasible, from which important epidemiological times, like the duration of the epidemic is derivable. This is an advantage of this graphical method that is illustrated below. It is clear that the line,

$$Y = \rho_{ccc}, \tag{21}$$

divides the graph into two regions; the upper region, which is the non-epidemic region and the lower region, which is the epidemic region.

$$Y = \frac{I(t)}{C(t)}, \tag{22}$$

gives, the case-carrier ratio trajectory over time. The points of intersection of equation (21) and (22) provide information about the duration times of the epidemic. Hence equation (21) and (22) are plotted over time. Here it is assumed the epidemic period is 90 days. The plot is extended to time $t = 225$ days; well over the epidemic period (and inside the rainy season) in order to provide visual evidence on how the case-carrier ratio trajectory vary over epidemiological context (endemicity, hyperendemicity and epidemic).

The plots are shown below for various parameter and starting values, for population size N=500 in Fig 6.1 and 6.2. The corresponding Susceptible, Carrier and Infective (SCI) plots for these parameters and starting values are also shown in Fig 6.3 and 6.4 so that they aide in the interpretation of the case-carrier ratio plots. Important epidemiological times and their corresponding numbers of susceptible, carrier and infectives identified from the case-carrier ratio profiles are shown in Tables 2 and 3. They aide in the understanding of the case-carrier ratio trajectories.

## 6.1 Discussion

It is observed that the case-carrier ratio trajectories (Fig 6.1 and 6.2) exhibit the same pattern, overtime, for all parameter, threshold ($\rho$), and initial values plotted except at the initial time period, where there is an observable difference. Trajectories with initial case-carrier ratio values greater that the threshold values, that is, in the non-epidemic region, experience a delay in time (denoted by $T_1$) before entering the epidemic region, where epidemic can start. Whereas for initial case-carrier ratio values less than the threshold value (that is, in the epidemic region), no such delay is observed. This supports the threshold condition for meningitis to spread derived earlier that meningitis will spread only when the initial case-carrier ratio is less than the threshold value. These highlighted differences in the values of $T_1$ are shown in Tables 2 and 3.

There are two major turning points observable in all the trajectories, the first one is a minima, its time of occurrence is denoted by $T_3$ and tabulated for all the trajectories in Tables 2 and 3. $T_3$, is the turning point at which the case-carrier ratio values changes from a decreasing to an increasing value. Hence it is indicative that at $T_3$ there is a critical case-carrier ratio value. In order to understand this value, we look at the SCI plots in Figs.6.3 and 6.4 at $T_3$. For the infective(I) curve, $T_3$ signifies the beginning of a sharp rising gradient which in turn indicates the beginning of rising incidence of infectives to an epidemic proportion. As shown in Tables 2.and 3, the value of the number of carriers is by far greater than that of the infectives at time $T_3$, suggesting that it is the carrier conversion rate that is propelling the rise in the incidence rate.

The rise in the incidence rate continues until the trajectory crosses the threshold line at time denoted by $T_2$, to enter into the non-epidemic region. In order to understand what is happening in the non-epidemic region better, again we look at the SCI plots. The time the infective curve attains its maximum value is denoted by $_IT_{max}$ and shown in Tables 2 and 3. Hence after $_IT_{max}$, there is a decreasing gradient in the infective curve, indicating a decreasing number of reported cases. Notice that the corresponding values of $T_2$ and $_IT_{max}$ in Tables 2 are the same. It is then clear that although above the threshold line is called a non-epidemic region, it is also noted that in this region, the number of reported cases begins to decline.

The second turning point occurs at time denoted by $T_4$, a maxima point. Here again the infective plots show that the numbers of reported cases continue to decline even after $T_4$. It is also observable that from time $T_4$, the number of infective becomes greater than the number of carriers, particularly for higher threshold values. From the SCI plots, by 50 days the epidemic is virtually over. After 90 days, well into the wet season, the epidemic has settled to an endemic level.

From the trajectory plots, it is clear that after 90 days, the case-carrier ratio remains constant and running parallel and close to their respective threshold lines. These suggest that the threshold value can be taking as an approximate estimate of the initial case-carrier ratio that will determine the start of the epidemic in the next dry season. The case-carrier ratio trajectory has provided pictorial evidence of how meningitis incidence varies according to epidemiological context, in this case, epidemic incidence, during the later part of the dry season and endemic incidence during the raining season. However, the introduction of the threshold condition has clearly further partitioned the epidemic incidence into periods of rising (below the threshold line) and declining (above the threshold line) epidemic, that can also be considered as hyper endemic.

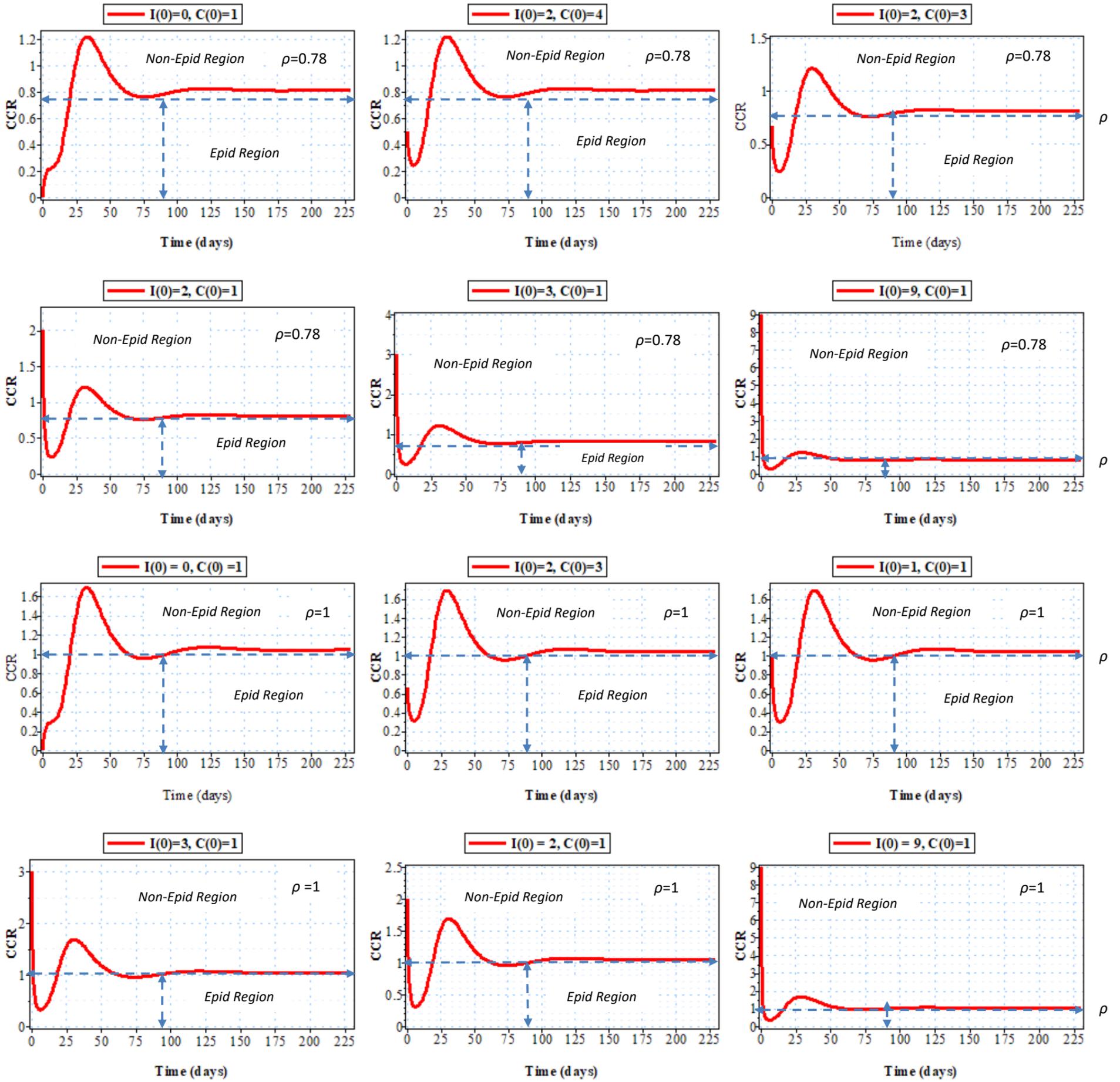

**Epid Region: Epidemic Region, Non-Epid Region: Non-Epidemic Region**

**Fig 6.1** Case –Carrier trajectory for various values of initial conditions and threshold values, $\rho = 0.78$ and $\rho = 1$, for population size N=500.

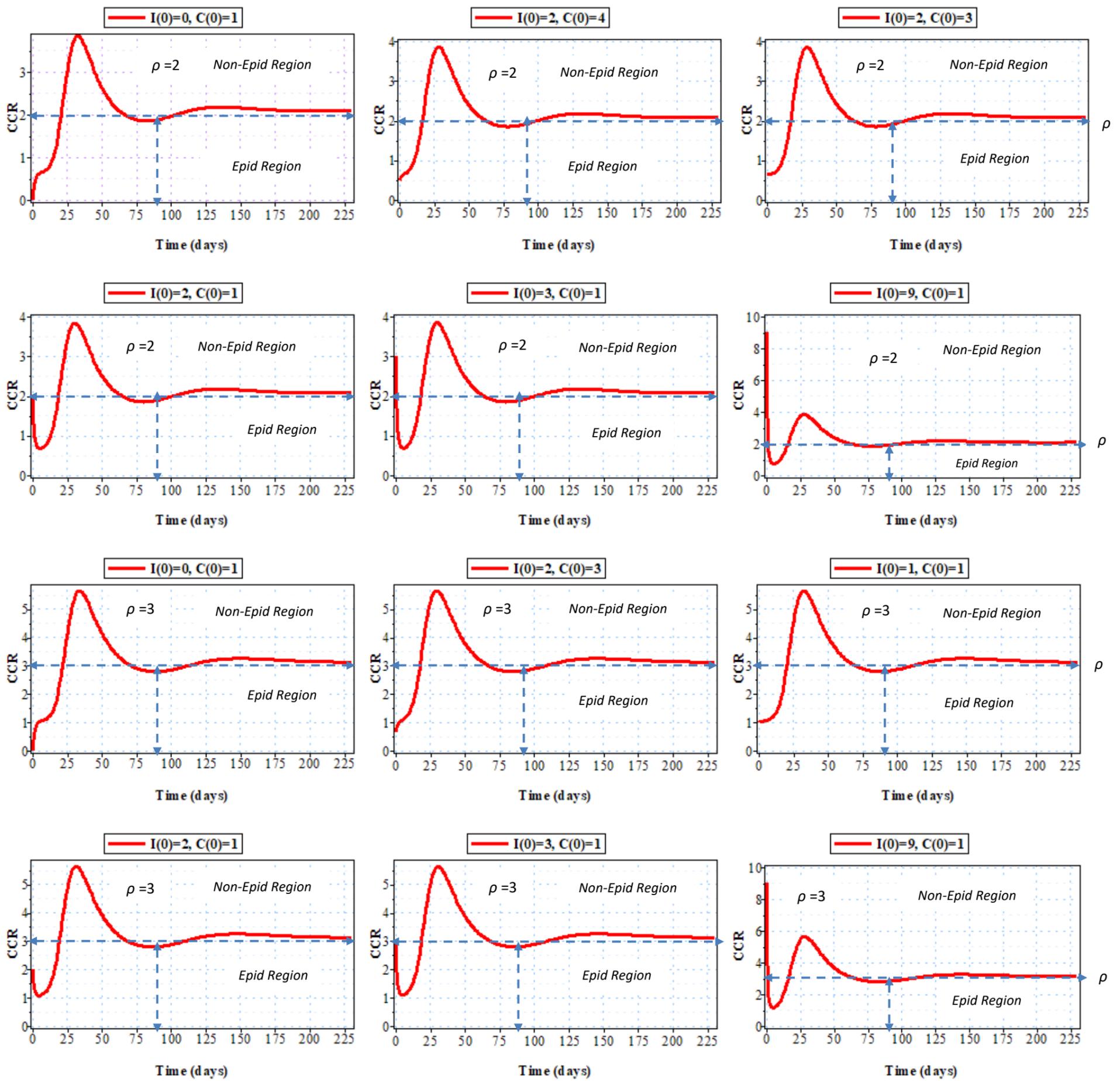

**Epid Region: Epidemic Region, Non-Epid Region: Non-Epidemic Region**

**Fig 6.2** Case –Carrier trajectory for various values of initial conditions and threshold values, $\rho = 2$ and $\rho = 3$, for population size N=500.

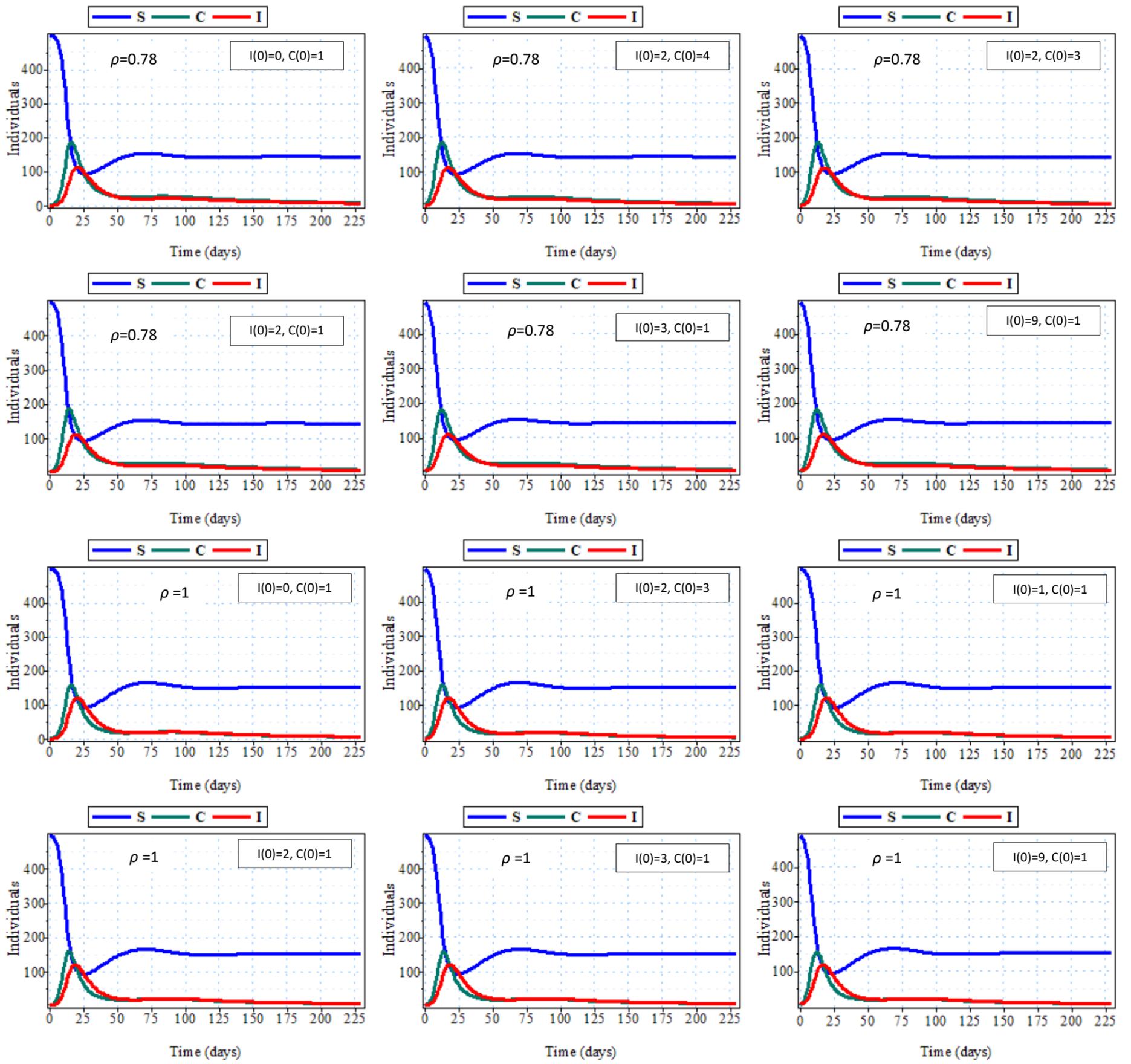

**Fig 6.3** The SCI plot for various values of initial conditions and threshold values, $\rho = 0.78$ and $\rho = 1$, for population size N=500.

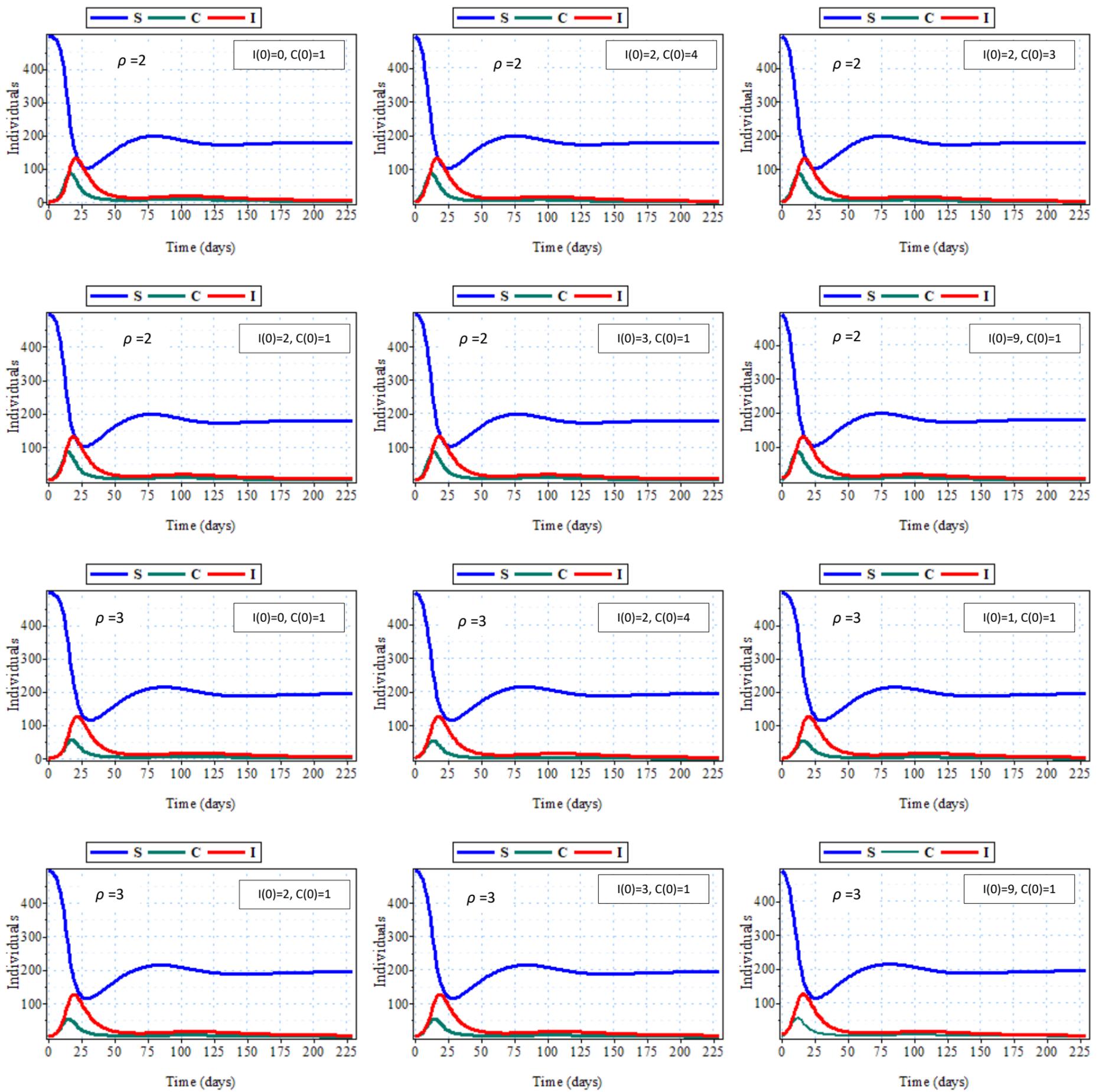

**Fig 6.4** The SCI plot for various values of initial conditions and threshold values, $\rho = 2$ and $\rho = 3$, for population size N=500.

**Table 2.:** Epidemiological characteristics of Meningitis Epidemic for $\rho = 0.78$ and $\rho = 1$, $N = 500$, and various initial conditions. All the epidemiological characteristics are in the unit of days.

| Epid Threshold ($\rho$) | $I(0)$ | $C(0)$ | Epid starting time ($T_1$) | End of Epid ($T_2$) | Duration of Epid ($T_2-T_1$) | $S(T_2)$ | $C(T_2)$ | $I(T_2)$ | Turning point time in Epid Region ($T_3$) | $S(T_3)$ | $C(T_3)$ | $I(T_3)$ | Turning point time in Non Epid Region ($T_4$) | $S(T_4)$ | $C(T_4)$ | $I(T_4)$ | $_cT_{max}$ | $S(_cT_{max})$ | $C(_cT_{max})$ | $I(_cT_{max})$ | $\rho$ | $_IT_{max}$ | $S(_IT_{max})$ | $C(_IT_{max})$ | $I(_IT_{max})$ |
|---|---|---|---|---|---|---|---|---|---|---|---|---|---|---|---|---|---|---|---|---|---|---|---|---|---|
| 0.78 | 0 | 1 | 0 | 21 | 21 | 104 | 141 | 111 | 3 | 493 | 5 | 1 | 35 | 104 | 43 | 51 | 16 | 173 | 186 | 84 | 0.78 | 21 | 104 | 141 | 111 |
| 0.78 | 2 | 4 | 0 | 17 | 17 | 105 | 143 | 111 | 5 | 436 | 47 | 11 | 33 | 107 | 40 | 48 | 13 | 163 | 186 | 88 | 0.78 | 17 | 105 | 143 | 111 |
| 0.78 | 2 | 3 | 0 | 18 | 18 | 103 | 139 | 111 | 5 | 426 | 54 | 13 | 35 | 106 | 39 | 47 | 13 | 171 | 186 | 85 | 0.78 | 18 | 103 | 139 | 111 |
| 0.78 | 2 | 1 | 1 | 19 | 18 | 104 | 142 | 111 | 6 | 466 | 39 | 8 | 34 | 104 | 43 | 51 | 14 | 177 | 185 | 82 | 0.78 | 19 | 104 | 142 | 111 |
| 0.78 | 3 | 1 | 2 | 20 | 18 | 103 | 140 | 110 | 6 | 452 | 37 | 9 | 35 | 105 | 42 | 50 | 14 | 172 | 185 | 84 | 0.78 | 20 | 103 | 140 | 110 |
| 0.78 | 9 | 1 | 3 | 17 | 14 | 104 | 139 | 109 | 7 | 404 | 66 | 18 | 33 | 107 | 39 | 47 | 12 | 176 | 183 | 82 | 0.78 | 17 | 104 | 139 | 109 |
| 1.00 | 0 | 1 | 0 | 20 | 20 | 110 | 119 | 120 | 5 | 488 | 8 | 2 | 36 | 108 | 28 | 46 | 16 | 181 | 159 | 94 | 1.00 | 20 | 110 | 119 | 120 |
| 1.00 | 2 | 3 | 0 | 17 | 17 | 109 | 117 | 120 | 5 | 440 | 22 | 10 | 33 | 106 | 28 | 47 | 13 | 175 | 158 | 97 | 1.00 | 17 | 109 | 117 | 120 |
| 1.00 | 1 | 1 | 0 | 20 | 20 | 110 | 118 | 120 | 7 | 459 | 28 | 8 | 34 | 105 | 30 | 49 | 15 | 178 | 158 | 96 | 1.00 | 20 | 110 | 118 | 120 |
| 1.00 | 2 | 1 | 1 | 19 | 18 | 113 | 122 | 120 | 7 | 474 | 17 | 5 | 33 | 105 | 30 | 51 | 14 | 188 | 158 | 91 | 1.00 | 19 | 113 | 122 | 120 |
| 1.00 | 3 | 1 | 2 | 18 | 16 | 111 | 120 | 120 | 6 | 455 | 30 | 10 | 35 | 111 | 26 | 41 | 13 | 183 | 158 | 93 | 1.00 | 18 | 111 | 120 | 120 |
| 1.00 | 9 | 1 | 3 | 16 | 13 | 113 | 121 | 119 | 5 | 438 | 40 | 14 | 32 | 107 | 28 | 46 | 12 | 188 | 156 | 90 | 1.00 | 16 | 113 | 121 | 119 |

**Epid:** Epidemic

**Table 2 continued.:** Epidemiological characteristics of Meningitis Epidemic for $\rho=2$ and $\rho=3$, N=500, and various initial conditions. All the epidemiological characteristics are in the unit of days.

| Epid Threshold ($\rho$) | $I(0)$ | $C(0)$ | Epid starting time ($T_1$) | End of Epid ($T_2$) | Duration of Epid ($T_2-T_1$) | $S(T_2)$ | $C(T_2)$ | $I(T_2)$ | Turning point time in Epid Region ($T_3$) | $S(T_3)$ | $C(T_3)$ | $I(T_3)$ | Turning point time in Non Epid Region ($T_4$) | $S(T_4)$ | $C(T_4)$ | $I(T_4)$ | $_cT_{max}$ | $S(_cT_{max})$ | $C(_cT_{max})$ | $I(_cT_{max})$ | $\rho$ | $_IT_{max}$ | $S(_IT_{max})$ | $C(_IT_{max})$ | $I(_IT_{max})$ |
|---|---|---|---|---|---|---|---|---|---|---|---|---|---|---|---|---|---|---|---|---|---|---|---|---|---|
| 2.00 | 0 | 1 | 0 | 20 | 20 | 146 | 67 | 131 | 5 | 488 | 6 | 4 | 35 | 109 | 14 | 52 | 17 | 228 | 87 | 105 | 2.00 | 20 | 146 | 67 | 131 |
| 2.00 | 2 | 4 | 0 | 16 | 16 | 145 | 67 | 131 | 4 | 472 | 15 | 10 | 33 | 115 | 11 | 41 | 12 | 229 | 87 | 106 | 2.00 | 16 | 145 | 67 | 131 |
| 2.00 | 2 | 3 | 0 | 17 | 17 | 145 | 67 | 131 | 4 | 467 | 17 | 11 | 30 | 110 | 13 | 49 | 13 | 228 | 87 | 106 | 2.00 | 17 | 145 | 67 | 131 |
| 2.00 | 2 | 1 | 1 | 19 | 18 | 141 | 64 | 131 | 5 | 474 | 13 | 9 | 35 | 121 | 11 | 43 | 14 | 240 | 86 | 101 | 2.00 | 19 | 141 | 64 | 131 |
| 2.00 | 3 | 1 | 2 | 18 | 16 | 141 | 64 | 131 | 5 | 476 | 12 | 8 | 33 | 112 | 12 | 47 | 14 | 238 | 86 | 101 | 2.00 | 18 | 141 | 64 | 131 |
| 2.00 | 9 | 1 | 3 | 16 | 13 | 146 | 66 | 130 | 5 | 443 | 27 | 19 | 30 | 110 | 13 | 49 | 12 | 228 | 86 | 105 | 2.00 | 16 | 146 | 66 | 130 |
| 3.00 | 0 | 1 | 0 | 22 | 22 | 171 | 42 | 126 | 4 | 494 | 3 | 2 | 38 | 123 | 7 | 41 | 18 | 268 | 54 | 99 | 3.00 | 22 | 171 | 42 | 126 |
| 3.00 | 2 | 3 | 0 | 17 | 17 | 169 | 42 | 126 | 4 | 472 | 9 | 10 | 34 | 124 | 7 | 41 | 13 | 265 | 55 | 101 | 3.00 | 17 | 169 | 42 | 126 |
| 3.00 | 1 | 1 | 0 | 20 | 20 | 171 | 42 | 126 | 6 | 482 | 7 | 7 | 35 | 120 | 8 | 47 | 16 | 269 | 54 | 99 | 3.00 | 20 | 171 | 42 | 126 |
| 3.00 | 2 | 1 | 0 | 19 | 19 | 169 | 42 | 126 | 5 | 477 | 8 | 9 | 36 | 124 | 7 | 41 | 15 | 265 | 54 | 100 | 3.00 | 19 | 169 | 42 | 126 |
| 3.00 | 3 | 1 | 1 | 18 | 17 | 172 | 42 | 126 | 6 | 459 | 15 | 17 | 36 | 125 | 7 | 40 | 14 | 269 | 54 | 98 | 3.00 | 18 | 172 | 42 | 126 |
| 3.00 | 9 | 1 | 2 | 17 | 15 | 171 | 42 | 126 | 5 | 446 | 19 | 22 | 31 | 121 | 8 | 46 | 12 | 265 | 54 | 100 | 3.00 | 17 | 171 | 42 | 126 |

**Epid:** Epidemic

# 7. Basic Reproductive Number

The basic reproductive number is the number of secondary infections produced by a single infectious person in a completely susceptible population. There are two approaches for determining the basic reproductive number: heuristic argument (Keeling and Rohani, 2007) and the next generation method (Diekmann et al., 2002, Van den Driessche and Watmough, 2002, and Diekmann et al., 2010). We use the method of Diekmann et al. (2010) just as have been used for epidemic process in Okunghae and Omame (2020). The states at infection of model 1, represented by the system of equation (2), are C and I. The linearized subsystem of these states at infection at disease free equilibrium point is given by

$$\frac{dC(t)}{dt} = (1-\varphi)\beta I + (1-\varphi)\epsilon\beta C - (\phi+\sigma)C \tag{23}$$

$$\frac{dI(t)}{dt} = \varphi\beta I + \varphi\epsilon\beta C + \phi C - (\theta+\gamma_1)I .$$

We partition equation (23) as

$$\frac{dV}{dt} = (T+W)V(t) \tag{24}$$

where

$$T = \begin{bmatrix} (1-\varphi)\epsilon\beta & (1-\varphi)\beta \\ \varphi\epsilon\beta & \varphi\beta \end{bmatrix}, \quad W = \begin{bmatrix} -(\phi+\sigma) & 0 \\ \phi & -(\theta+\gamma_1) \end{bmatrix} \text{ and } V = \begin{bmatrix} C \\ I \end{bmatrix}.$$

Using the method in Diekmann et al.(2010), the next generation matrix denoted by

$$G = -TW^{-1}$$

$$= \begin{bmatrix} \dfrac{(1-\varphi)\epsilon\beta}{(\phi+\sigma)} + \dfrac{(1-\varphi)\beta}{(\phi+\sigma)(\theta+\gamma_1)} & \dfrac{(1-\varphi)\beta}{(\theta+\gamma_1)} \\ \dfrac{\varphi\epsilon\beta}{(\phi+\sigma)} + \dfrac{\varphi\beta\phi}{(\phi+\sigma)(\theta+\gamma_1)} & \dfrac{\varphi\beta}{(\theta+\gamma_1)} \end{bmatrix} \tag{25}$$

The dominant eigenvalue of the $2 \times 2$ matrix $G$ gives the basic reproductive number $R_0^{G_1}$, which is of interest here. This can be obtained from the equation of the spectral radius of $G$ given as

$$R_0^{G_1} = \frac{Trace(G) + \sqrt{Trace(G)^2 - 4\det(G)}}{2}$$

$$= \frac{(1-\varphi)\epsilon\beta}{(\phi+\sigma)} + \frac{(1-\varphi)\beta}{(\phi+\sigma)(\theta+\gamma_1)} + \frac{\varphi\beta}{(\theta+\gamma_1)}. \tag{26}$$

Note that $R_0^{G_1}$ comprises of three components indicating the different sources of contribution to the reproductive number. We define these below so as to give better understanding of the role played by each, in meningitis epidemic process. Let

$$_cR_0^{G_1} = \frac{(1-\varphi)\epsilon\beta}{(\phi+\sigma)}, \quad _{cI}R_0^{G_1} = \frac{(1-\varphi)\beta}{(\phi+\sigma)(\theta+\gamma_1)} \quad \text{and} \quad _IR_0^{G_1} = \frac{\varphi\beta}{(\theta+\gamma_1)}.$$

Note that $_cR_0^{G_1}$ is the number of secondary infective produced by a carrier during average time of duration of carriership and $_IR_0^{G_1}$, is the number of secondary infective individuals that become infective directly, without passing through carriership, produced by a symptomatic infective.

**Note 1.** $_cR_0^{G_1}$, is the same as $_cR_0$ given in equation (12) and derived as the condition for spread of carrier infectives when the initial number of infectives, I(0), is zero.

**Note 2.** $_IR_0^{G_1}$, is the same as $_IR_0$ given in equation (15), derived as the condition for the spread of symptomatic meningitis when the initial number of carrier is zero.

These notes have provided a greater understanding of the role of $R_0^{G_1}$ in the dynamics of meningitis epidemic process. Consequently, we can draw conclusion reasoning as earlier, that for model 1, any control intervention targeted at reducing the proportion of susceptibles below $1/R_0^{G_1}$ will eradicate the cases of meningitis.

Model 2 is a special case of model 1 when $\varphi = 0$. Hence

$$R_0^{G_2} = {}_cR_0^{G_2} + {}_{CI}R_0^{G_2} \tag{27}$$

where

$${}_cR_0^{G_2} = \frac{\epsilon\beta}{(\phi+\sigma)}, \quad {}_{CI}R_0^{G_2} = \frac{\phi\beta}{(\phi+\sigma)(\theta+\gamma_1)}$$

Here again, just like before ${}_cR_0^{G_2}$ is the number asymptomatic infective produced by carrier while ${}_{CI}R_0^{G_2}$ is the number of secondary symptomatic infective passing through carriership that are produced by a symptomatic infective. By similar reasoning as before, for model 2, any control intervention targeted at reducing the proportion of susceptibles below $1/R_0^{G_2}$ will eradicate the cases of meningitis.

## 8. Sensitivity Analysis

Here interest lies in computing the sensitivity indices of $R_0^{G_1}$ and $R_0^{G_2}$ to changes in model parameters. These indices provide relative importance of model parameters in the meningitis transmission process. Consequently they serve as crucial information for any control strategies. The normalized forward sensitivity index (Chitnis et al., 2008) of a variable to a parameter as defined in Ndelwa et al. (2015) and stated below is used for the computation. The sensitivity of the reproductive number, $R_0^{G_i}$, for example $\beta$, is denoted by

$$S_\beta = \frac{\partial R_0^{G_i}}{\partial \beta}\left(\frac{\beta}{R_0^{G_i}}\right). \qquad i = 1, 2 \tag{16}$$

These indices are computed for models 1 and 2 at specified parameter values as shown in the results in tables 6 and 7 respectively. Indices for the components that make up $R_0^{G_i}(i = 1,2)$ are also computed in order to shed more light on the dynamics of the infection process.

## 8.1 Discussion

The results in Table 6 show that only indices for $\beta$ and $\epsilon$ are positive. That is, when the parameters $\beta$ and $\epsilon$ are increased while keeping the values of the other parameters constant, $R_0^{G_1}$ will increase. This increase suggests that they have positive impact on the reproductive number indicating that they are possible targets for any control intervention strategy. The indices for other parameters $\sigma, \phi, \gamma_1$ and $\theta$, are negative, consequently, they are adjudged to be of less concern for any control intervention.

The results for the components of $R_0^{G_1}, {}_C R_0^{G_1}, {}_{CI} R_0^{G_1}$ and ${}_I R_0^{G_1}$ also identify β and ϵ as targets for control intervention. In addition $\phi$ is identified in ${}_{CI} R_0^{G_1}$ as having positive impact. This could be an additional target for control intervention. This is reasonable as $\phi$ defines the rate of conversion of carriers to symptomatic infectives, the reduction of which will slow down the dynamics of the infection process. The conclusions for model 2 from Table 7 are similar as in model 1. The parameters $\beta, \epsilon$ and $\phi$ are also identified as important parameters for any control intervention.

.

**Table 6**: Sensitivity analysis for parameters of the basic reproductive number for model 1

| | | | | | Sensitivity | | | | | | | | | |
|---|---|---|---|---|---|---|---|---|---|---|---|---|---|---|
| Model 1 | | $\varepsilon = 2$ $\varphi = 0.5$ | | | | | | | $\phi = 0.14247$ $\varphi = 0.5$ | | | $\phi = 0.14247$ $\varepsilon = 2$ | | |
| Parameter | $\beta$ | $\phi$ | | | | $\sigma$ | $\theta$ | $\gamma_1$ | $\varepsilon$ | | | $\varphi$ | | |
| Values | 0.34247 | 0.14247 | 0.18333 | 0.36666 | 0.54999 | 0.14247 | 0.08333 | 0.1 | 0.5 | 1 | 2 | 0.2 | 0.5 | 0.8 |
| $_C R_0^{G_1}$ | 1 | -0.5 | -0.56271 | -0.72017 | -0.79426 | -0.5 | 0 | 0 | 1 | 1 | 1 | -0.25 | 1 | 4 |
| $_{CI} R_0^{G_1}$ | 1 | 0.5 | 0.43729 | 0.27983 | 0.20574 | -0.5 | -0.45453 | -0.5455 | 0 | 0 | 0 | -0.25 | 1 | 4 |
| $_I R_0^{G_1}$ | 1 | 0 | 0 | 0 | 0 | 0 | -0.45453 | -0.5455 | 0 | 0 | 0 | 1 | 1 | 1 |
| $R_0^{G_1}$ | 1 | 0.1412 | 0.1440 | 0.1299 | 0.1107 | -0.321 | -0.24465 | -0.2936 | 0.177 | 0.30 | 0.462 | -0.39 | -0.28 | -0.14 |

**Table 7**: Sensitivity analysis for parameters of the basic reproductive number for model 2

| | | | | | Sensitivity | | | | | | |
|---|---|---|---|---|---|---|---|---|---|---|---|
| Model 2 | | $\varepsilon = 2$ | | | | | | | $\phi = 0.14247$ | | |
| Parameter | $\beta$ | $\phi$ | | | | $\sigma$ | $\theta$ | $\gamma_1$ | $\varepsilon$ | | |
| Values | 0.34247 | 0.14247 | 0.18333 | 0.36666 | 0.54999 | 0.14247 | 0.08333 | 0.1 | 0.5 | 1 | 2 |
| $_C R_0^{G_2}$ | 1 | -0.5 | -0.56271 | -0.72017 | -0.79426 | -0.5 | 0 | 0 | 1 | 1 | 1 |
| $_{CI} R_0^{G_2}$ | 1 | 0.5 | 0.43729 | 0.27983 | 0.20574 | -0.5 | -0.45453 | -0.54546 | 0 | 0 | 0 |
| $R_0^{G_2}$ | 1 | -0.22017 | -0.22937 | -0.22017 | -0.19426 | -0.5 | -0.12719 | -0.15264 | 0.39150 | 0.56271 | 0.72017 |

The results in Table 8 show that indices are positive and increase with increasing $\epsilon$ for all the given values of $\varphi$, whereas in Table 9, the values of the indices are negative for a stipulated $\epsilon$ greater than one, showing a decreasing trend for increasing $\varphi$ for each of these stipulated $\epsilon$. Consequently, $\epsilon$ is a more important variable choice for any control intervention than $\varphi$.

**Table 8**. Effect of $\epsilon$ on $R_0^{G_1}$ for $\varphi = 0.0, 0.5$ and $0.8$; $\beta = 0.34247; \phi = 0.14247;$ $\sigma = 0.14247; \theta = 0.08333$ and $\gamma_1 = 0.1$

| | Sensitivity ($R_0^{G_1}$) | | |
|---|---|---|---|
| | | $\varphi$ | |
| $\epsilon$ | 0.0 | 0.5 | 0.8 |
| 0.1 | 0.114 | 0.041 | 0.014 |
| 0.5 | 0.392 | 0.177 | 0.066 |
| 1.0 | 0.563 | 0.300 | 0.125 |
| 1.5 | 0.659 | 0.392 | 0.177 |
| 2.0 | 0.720 | 0.462 | 0.222 |
| 2.5 | 0.763 | 0.517 | 0.263 |
| 3.0 | 0.794 | 0.563 | 0.300 |
| 3.5 | 0.818 | 0.600 | 0.333 |
| 4.0 | 0.837 | 0.632 | 0.364 |

**Table 9**. Effect of $\varphi$ on $R_0^{G_1}$ for $\epsilon = 0.5, 1$ and $2$; $\beta = 0.34247; \phi = 0.14247;$ $\sigma = 0.14247; \theta = 0.08333$ and $\gamma_1 = 0.1$

| | Sensitivity ($R_0^{G_1}$) | | |
|---|---|---|---|
| | | $\epsilon$ | |
| $\varphi$ | 0.5 | 1 | 2 |
| 0.0 | 0.000 | 0.000 | 0.000 |
| 0.1 | 0.021 | -0.013 | -0.046 |
| 0.2 | 0.042 | -0.026 | -0.097 |
| 0.3 | 0.061 | -0.039 | -0.152 |
| 0.4 | 0.080 | -0.053 | -0.214 |
| 0.5 | 0.100 | -0.067 | -0.282 |
| 0.6 | 0.115 | -0.081 | -0.359 |
| 0.7 | 0.132 | -0.096 | -0.446 |
| 0.8 | 0.148 | -0.115 | -0.544 |

## 9. Effect of $\epsilon$ and $\varphi$ on some meningitis epidemic indicators

Here focus is on the effect of $\epsilon$ and $\varphi$ on the value of the reproductive number, $R_0^{G_1}$ and the total proportion ever infected during the epidemic duration which we assume to be 90 days. This is done in order to gain further insight into the meningitis infection process. We first computed the values of $R_0^{G_1}$ for increasing values of $\varphi$ for $\epsilon = 0.5, 1$ and 2 (Table 10) and the values of $R_0^{G_1}$ for increasing values of $\epsilon$ for $\varphi = 0, 0.5$ and 0.8 (Table 11).

It is clear from Table 10 and 11 that the lowest index value for $R_0^{G_1}$, is achieved where $\epsilon = 0.1$ and $\varphi = 0$ as seen in table 11. This implies that, $\epsilon$, is a more important target intervention parameter than $\varphi$. Consequently, a control intervention strategy focusing on reducing $\epsilon$ can be used in achieving a desirable level of value of $R_0^{G_1}$.

**Table 10**. Effect of $\varphi$ on $R_0^{G_1}$ for $\epsilon = 0.5, 1$ and 2; $\beta = 0.34247; \phi = 0.14247;$
$\sigma = 0.14247; \theta = 0.08333$ and $\gamma_1 = 0.1$

| | $R_0^{G_1}$ | | |
|---|---|---|---|
| | $\epsilon$ | | |
| $\varphi$ | **0.0** | **0.5** | **0.8** |
| 0.0 | 0.632 | 1.233 | 2.435 |
| 0.1 | 0.755 | 1.297 | 2.378 |
| 0.2 | 0.880 | 1.360 | 2.322 |
| 0.3 | 1.000 | 1.424 | 2.265 |
| 0.4 | 1.266 | 1.487 | 2.208 |
| 0.5 | 1.250 | 1.551 | 2.152 |
| 0.6 | 1.374 | 1.614 | 2.095 |
| 0.7 | 1.497 | 1.678 | 2.038 |
| 0.8 | 1.621 | 1.741 | 1.981 |

**Table 11**. Effect of $\epsilon$ on $R_0^{G_1}$ for $\varphi = 0.0, 0.5$ and 0.8; $\beta = 0.34247; \phi = 0.14247;$
$\sigma = 0.14247; \theta = 0.08333$ and $\gamma_1 = 0.1$

| | $R_0^{G_1}$ | | |
|---|---|---|---|
| | $\varphi$ | | |
| $\epsilon$ | 0.0 | 0.5 | 0.8 |
| 0.1 | 0.152 | 1.000 | 1.525 |
| 0.5 | 0.632 | 1.250 | 1.621 |
| 1.0 | 1.233 | 1.551 | 1.741 |
| 1.5 | 1.834 | 1.851 | 1.861 |
| 2.0 | 2.435 | 2.152 | 1.981 |
| 2.5 | 3.036 | 2.452 | 2.101 |
| 3.0 | 3.637 | 2.753 | 2.222 |
| 3.5 | 4.238 | 3.053 | 2.342 |
| 4.0 | 4.840 | 3.354 | 2.462 |

The effect of $\varphi$ and $\epsilon$ on the total proportion ever infected using graphs were investigated. The results are presented in figures 9.1a and b below. Here again, the minimum total proportion ever infected is achieved when $\epsilon = 0.1$ and $\varphi = 0$. Hence the conclusion here is same as drawn from Tables 10 and 11, that is, $\epsilon$ is a very important parameter for every control intervention strategy .It should be noted that since $\varphi = 0$ in model 2, consequently model 2 can serve as a good first approximation to model 1 that can be used to represent the salient features of meningitis infection process.

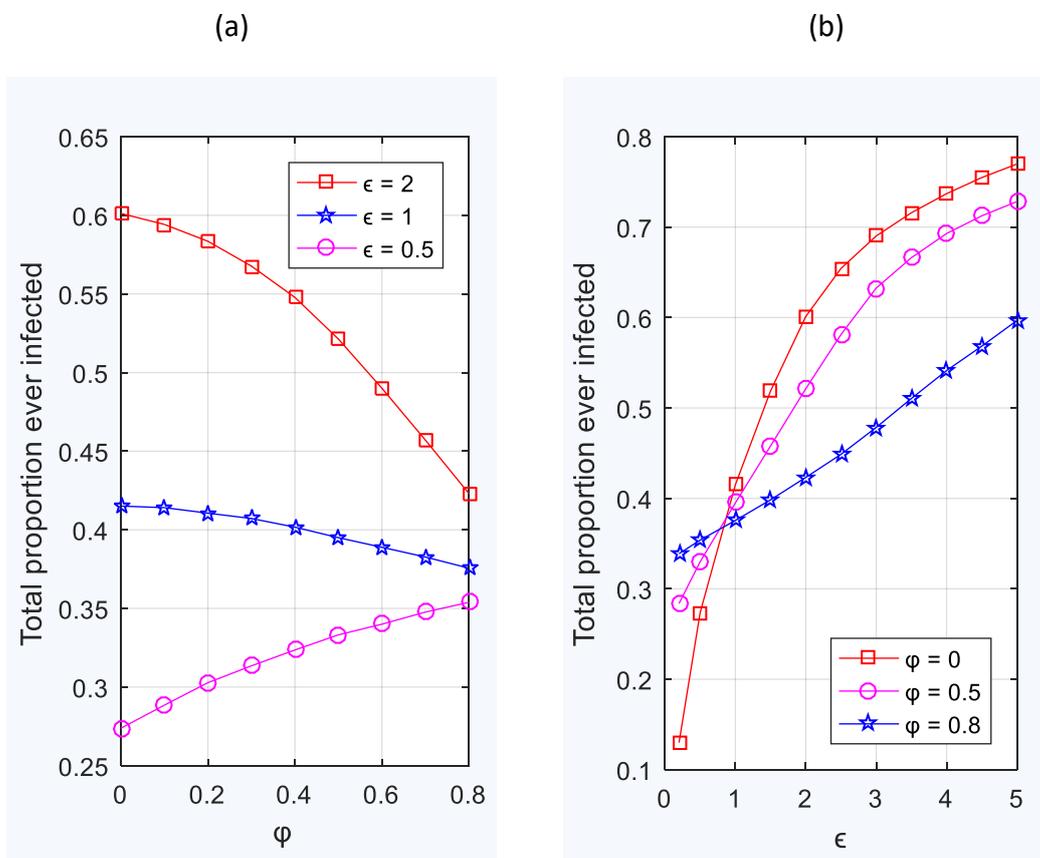

**Fig 9 .1**. Plot of total proportion ever infected at the end of 90 days, for initial condition I(0) = 1, C(0) = 1 and (a) Varying $\varphi$ for $\epsilon = 0.5, 1, 2$ (b) Varying $\epsilon$ for $\varphi = 0, 0.5, 0.8$

## 10. Conclusion

A $SCII_1RS^{CII_1}$ model that allows immunity from both stages of infection and carriage and disease induced death is used to describe the salient features of meningitis epidemic in a closed population. This is new. It allows the odds, $\epsilon$. in favor of a carrier, transmitting infection to a susceptible to be different from one. The model is generalized (denoted by $SCII_1RS^{ALT}$) to allow a proportion ($\varphi$) of those infected progress directly from S to I. The threshold conditions suggests that meningitis can be eradicated if the initial proportion of susceptible is brought below the inverse of whichever is greater of the two reproductive numbers due to carrier or an infective obtained in the generalized model.

The parameters $\beta, \epsilon$ and $\varphi$ are identified as important targets that need be reduced for any meaningful attempt at controlling the spread of meningitis. The case-carrier ratio profiles indicate there is a critical case-carrier ratio is attained before the incidence of meningitis can enter an epidemic dimension. It also provides visual evidence of epidemiological context, that is, epidemic incidence in the later part of the dry season and endemic incidence during the rains. The graphs of the total proportion ever infected suggest that the model with $\varphi = 0$, that is, $SCII_1RS^{CII_1}$, can adequately replace the general model, where $\varphi \neq 0$, in describing the salient features of meningitis, that can be used in studying the dynamics of its transmission process.